\documentclass[10pt,letterpaper,oneside]{article}

\usepackage[
    textwidth=404.295pt,
    textheight=661.32pt,
    hcentering,
    top=60pt,
    headheight=0pt,
    headsep=0pt,
    footskip=30pt
]{geometry}

\usepackage{amsmath}
\usepackage{newtxtext}
\usepackage{newtxmath}
\usepackage{microtype}
\usepackage{graphicx,xcolor}
\usepackage{caption}
\usepackage[hidelinks]{hyperref}
\usepackage{apacite}
\AtBeginDocument{%
    
    \renewcommand{\BRetrieved}[1]{}
}

\let\cite\shortcite

\makeatletter
\renewcommand\normalsize{%
    \@setfontsize\normalsize{9}{12}%
    \abovedisplayskip=8.5pt plus 3pt minus 4pt
    \belowdisplayskip=8.5pt plus 3pt minus 4pt
    \abovedisplayshortskip=0pt plus 2pt
    \belowdisplayshortskip=4pt plus 2pt minus 2pt
    \let\@listi\@listI
}
\renewcommand\small{\@setfontsize\small{8}{10}}
\renewcommand\footnotesize{\@setfontsize\footnotesize{7}{9}}
\renewcommand\scriptsize{\@setfontsize\scriptsize{7}{8}}
\renewcommand\tiny{\@setfontsize\tiny{5}{6}}
\renewcommand\large{\@setfontsize\large{11}{12}}
\renewcommand\Large{\@setfontsize\Large{12}{12}}
\renewcommand\LARGE{\@setfontsize\LARGE{14}{16}}
\renewcommand\huge{\@setfontsize\huge{15}{18}}
\normalsize

\renewcommand\section{\@startsection{section}{1}{\z@}%
    {-20pt plus -2pt minus -2pt}{3pt plus 1pt}%
    {\normalfont\large\bfseries\raggedright}}
\renewcommand\subsection{\@startsection{subsection}{2}{\z@}%
    {-14pt plus -2pt minus -2pt}{2pt plus 1pt}%
    {\normalfont\normalsize\bfseries\raggedright}}
\renewcommand\subsubsection{\@startsection{subsubsection}{3}{\z@}%
    {-12pt plus -2pt minus -2pt}{2pt plus 1pt}%
    {\normalfont\normalsize\bfseries\raggedright}}

\renewcommand\maketitle{%
    \begingroup
    \centering
    {\huge\bfseries\@title\par}
    \vspace{8pt}
    {\fontsize{9}{10.8}\selectfont\@author\par}
    \vspace{3pt}
    {\small\paperaffiliation\par}
    \vspace{3pt}
    {\footnotesize\papernote\par}
    \endgroup
    \vspace{12pt}
}
\makeatother

\renewenvironment{abstract}
    {\par\begingroup\normalsize\noindent\textbf{Abstract.}\enspace\ignorespaces}
    {\par\endgroup\vspace{3pt}}
\newenvironment{plainlanguagesummary}
    {\par\begingroup\normalsize\noindent\textbf{Plain-language summary.}\enspace\ignorespaces}
    {\par\endgroup\vspace{3pt}}

\title{A more predictable Madden-Julian Oscillation index derived from Koopman spectral analysis}
\author{\textbf{Claire Valva}\textsuperscript{1,*},
        \textbf{Edwin P. Gerber}\textsuperscript{1}}
\newcommand{\paperaffiliation}{%
    \textsuperscript{1}Courant Institute School of Mathematics, Computing, and Data Science,
    New York University}
\newcommand{\papernote}{%
    \textsuperscript{*}Now Computing and Mathematical Sciences,
    California Institute of Technology.\par
    Correspondence: \href{mailto:clairev@caltech.edu}{clairev@caltech.edu}}
\date{}

\begin{document}
\maketitle

\begin{abstract}

    The Madden-Julian oscillation (MJO) is a major source of subseasonal-to-seasonal (S2S) predictability. The MJO is commonly defined and tracked with indices such as the Real-time Multivariate MJO (RMM) index. Although the RMM provides a useful description of the MJO, its evolution can be noisy and difficult to predict. We define an MJO index using a data-driven approximation of the Koopman operator. The Koopman index captures similar tropical circulation and convection patterns to the RMM but evolves more smoothly and predictably. Skillful prediction extends to 46 days for the Koopman index compared to 11 days for the RMM under the same prediction framework.  While this new approach does not recover the RMM as well as operational S2S models, which provide skillful forecasts up to 35 days, the Koopman index could complement existing MJO diagnostics in evaluating and developing extended-range forecast systems.

     \begin{plainlanguagesummary}
            The Madden-Julian oscillation (MJO) is an eastward moving pattern of enhanced and suppressed tropical rainfall that lasts, approximately, from 30 to 60 days. The MJO is connected with many aspects of global weather including tropical cyclones and the monsoon. Here, we use a data-driven method to identify an index of the MJO and show that this is a more predictable index than those traditionally used.
     \end{plainlanguagesummary}

\end{abstract}

\section{Introduction}  
The Madden-Julian oscillation (MJO) is the most prominent mode of intraseasonal variability in the tropical atmosphere. It is characterized by large-scale coupled patterns in atmospheric circulation and deep convection that propagate eastward with a phase speed of $5 \, \text{ms}^{-1}$ and a lifespan of 30-60 days \cite{maddenDetection4050Day1971, zhangMaddenJulianOscillation2005}. The MJO is known to influence many aspects of global climate and weather, including the onset of global monsoons and tropical cyclones, as well as extreme temperature and precipitation events \cite{jiangFiftyYearsResearch2020, jonesGlobalOccurrencesExtreme2004}. Due to its wide-ranging impacts on global weather and its quasi-periodic occurrence, the MJO is one of the primary sources of predictability for subseasonal-to-seasonal (S2S) forecasts \cite{vitartSubseasonalSeasonalS2S2017a}.

The MJO is often defined and tracked by indices based on winds, outgoing longwave radiation (OLR), or other convection proxies. The Real-time Multivariate MJO (RMM) index, initially derived in \citeA{wheelerAllSeasonRealTimeMultivariate2004}, is the most used index for tracking the MJO \cite{kiladisComparisonOLRCirculationBased2014, straubMJOInitiationRealTime2013}. The RMM is based on an empirical orthogonal function (EOF) analysis, defining a 2D phase space where the index propagates counterclockwise, from which the amplitude and the phase are computed. While this index is widely used, its propagation is relatively noisy, making it more challenging to predict.

Koopman operator theory maps finite-dimensional nonlinear dynamical systems to linear, albeit infinite dimensional, systems \cite{koopmanHamiltonianSystemsTransformation1931}. Eigendecompositions of the Koopman operator can be used to identify dynamically meaningful oscillatory modes with associated evolution frequencies \cite{mezicSpectralPropertiesDynamical2005a}. Such modes have been used to analyze and predict elements of the climate system with particular success for oscillatory phenomena. This was done initially for the MJO in \citeA{lintnerIdentificationMaddenJulian2023}, but has also had success in El Ni\~no Southern Oscillation (ENSO) and the Quasi-Biennial Oscillation \cite{wangExtendedrangeStatisticalENSO2020, froylandSpectralAnalysisClimate2021, valvaQBOAnnualCycle2025, lorenzo-sanchezKoopmanTheoryEnhanced2025, lorenzo-sanchezResidualPseudospectraReveal2026}.

Here, we define an alternative index of the MJO using a data-driven approximation of the Koopman operator. While the Koopman index and the RMM provide similar descriptions of the MJO when viewed through conditional statistics, the Koopman based index evolves more smoothly and can be predicted skillfully on intraseasonal timescales.

\section{Extracting intraseasonal oscillations from data with Koopman operator methods}
\label{sec:theory data}

We provide an overview of Koopman operator theory, and how to use it to extract oscillatory modes from tropical winds and OLR. Consider a nonlinear dynamical system with states $\{x_s\}_{s \in \mathbb{R}}$ --- in our case the tropical atmosphere --- and a flow map $\varphi^t$ so that $\varphi^t(x_s) = x_{s + t}$. The map $\varphi^t$ can be thought of as a model that integrates forward the state of the atmosphere from time $s$ to time $s + t$. The \textit{Koopman operator} $K^t$ acts on observables $f$ (functions) of the state
\begin{equation}
    K^t f (x_s) = f \circ \varphi^t (x_s) = f(x_{s + t}). 
\end{equation}
An example observable $f$ is the zonal wind speed at a given location above the Indian Ocean. Then $f(x_s)$ will give the zonal wind speed measurement at that location at time $s$, and $(K^tf)(x_s)$ the wind speed at time $s + t$.

While the dynamics $\varphi^t$ are nonlinear, the Koopman operator $K^t$ is linear when viewed as an operator on a space of observables $f$. As such, the advantage of the Koopman operator formalism is that we can perform a spectral analysis of $K^t$, which can be interpreted as coherent feature extraction. If $(\omega_k, \zeta_k)$ is an eigenfrequency-eigenfunction pair of $K^t$, then
\begin{equation}
    K^t \zeta_k = e^{t \omega_k} \zeta_k.
\end{equation}
The eigenfunctions (the coherent features) $\zeta_k$ evolve predictably, oscillating with frequency $\omega_k$. 

We can associate a spatio-temporally varying mode $M_k$ to the eigenpair $(\omega_k, \zeta_k)$ when we approximate the operator from data. This mode $M_k$ is computed from a projection of the input data onto the eigenfunction $\zeta_k$ and represents the variability of the projected variables associated to that eigenfunction. We will identify a particular Koopman mode of variability $M_k$ with the MJO. Then the eigenfunction $\zeta_k$ can serve as an MJO index.

We construct the approximate Koopman operator with ERA5 \cite{hersbachERA5GlobalReanalysis2020} outgoing longwave radiation (OLR) and zonal winds at $200$ and $850 \, \text{hPa}$ at $1.5^\circ$ spatial resolution, meridionally averaged from $15^\circ$ N to $15^\circ$ S. We normalize the OLR and zonal wind fields to be mean zero with unit variance in each variable, following the RMM computation convention \cite{wheelerAllSeasonRealTimeMultivariate2004} and previously defined Koopman MJO index \cite{lintnerIdentificationMaddenJulian2023}. The modes are defined using data from January 2000 to December 2019, and the later forecasting experiment will be validated on the period from December 2020 to December 2024. We additionally \textit{delay-embed} the data with an embedding window of $N_e = 64 \, \mathrm{days}$ before analysis. Delay embedding replaces the vector of input data $\mathbf{X}_j$ at a given time indexed as $j$ with $\hat{\mathbf{X}}_j$ such that the $j$th time and the $N_e$ previous times are included in that snapshot, i.e., $\hat{\mathbf{X}}_j = \{ \mathbf{X}_j, \mathbf{X}_{j - 1}, \dotsc, \mathbf{X}_{j - N_e} \}$. This step serves to insert ``memory'' and uniqueness to each state, a relevant step given the compressed nature of the meridionally averaged dataset \cite{ghilAdvancedSpectralMethods2002a}. We will refer to this final data array as $\mathbf{X} \in \mathbb{R}^{N_t \times N_d}$ where $N_t$ is the number of delay embedded snapshots and $N_d$ is the dimension of each snapshot.

We use the Koopman approximation algorithm described in \citeA{giannakisDatadrivenSpectralDecomposition2019}. This algorithm assumes the dynamical system of interest has an invariant probability measure (a stationary climatology), resulting in the eigenfrequencies $\omega_k$ being entirely imaginary. The approximate Koopman operator can then be written as

\begin{equation}
    K^t f \approx \sum_k e^{t\omega_k} \zeta_k c_k, \: c_k = \langle \zeta_k, f \rangle.
\end{equation}
The original dataset can be reconstructed from the corresponding Koopman modes $M_k \in \mathbb{C}^{N_t \times N_d}$,
\begin{equation}
    \mathbf{X} \approx \sum_{k} M_k. 
\end{equation}
The goal is to find a small number of modes $M_k$ that capture variability of interest.

\begin{figure*}[ht!]
    \centering
    \includegraphics[width=1\textwidth]{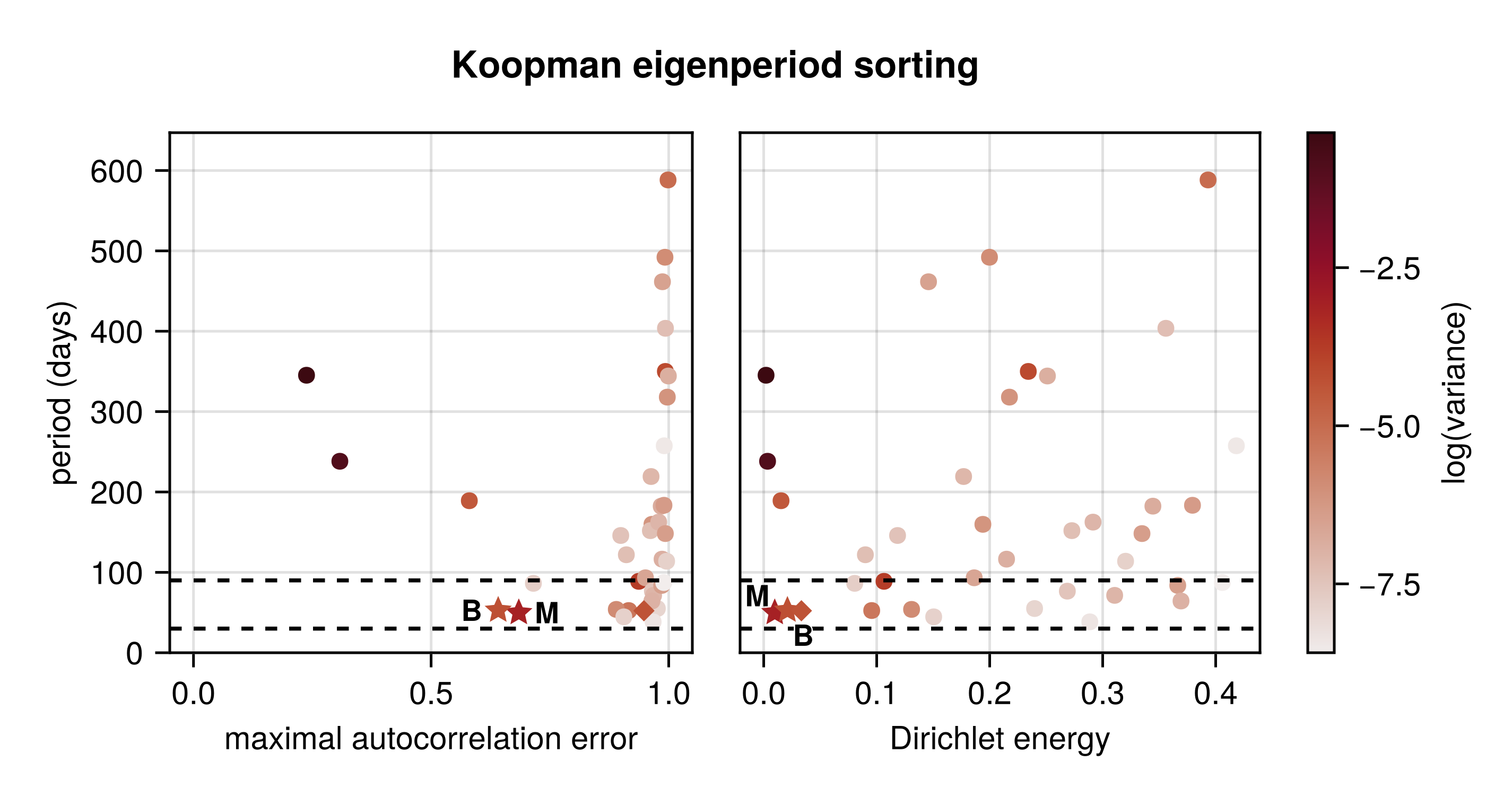}
    \caption{Koopman eigenperiods plotted according to two sorting metrics. The left panel sorts by a maximal autocorrelation error which ranges from 0 (best) to 1 (worst), while the right panel sorts by the Dirichlet energy metric where a smaller absolute value indicates a better, less noisy mode. The y-axis indicates the eigenperiod of each mode, while the x-axis is the value computed by each sorting metric. Marker color is associated with the variance of the mode. The horizontal lines indicate the period cutoff for intraseasonal modes (30-90 days). The stars indicate two intraseasonal modes of interest, interpreted to be the MJO (approx. $50\, \mathrm{days}$ period, labeled \textbf{M})  and the BSISO (approx. $53 \, \mathrm{days}$ period, labeled \textbf{B}). The two points outside of the intraseasonal band that remain in good sorting positions are likely related to the annual cycle.}
    \label{fig:eigenvalues}
\end{figure*}

To use the resulting approximation to extract a dynamically meaningful representation of the MJO from data, we must address two questions. First, can we find a Koopman mode with the right frequency and spatiotemporal variability of the MJO? Second, is the resulting Koopman eigenpair (and associated Koopman mode) robust?

To begin answering these questions, we plot the periods of the obtained Koopman modes in figure \ref{fig:eigenvalues}. We can first target an eigenmode through its frequency $\omega_k$. For example, to identify a representation of the annual cycle, we choose the eigenpair with $\omega_k = 1 \text{yr}^{-1}$ and potentially higher harmonics. In the figure, we see a harmonic of the annual cycle around $\omega_k = 2 \text{yr}^{-1}$ with a period of about 190 days. For the MJO, we will search for modes with an intraseasonal period, 30-90 days (dashed lines). There are many Koopman modes with periods within this intraseasonal range, but most of these eigenvalues represent a very small amount of variability (marker color) and  so are insignificant. 

To differentiate between the eigenpairs with the correct timescale of variability and meaningful variance, we assess their ``trustworthiness'' with two sorting metrics. First, we employ a metric of predictability called \textit{maximal autocorrelation error} \cite<proposed in>{giannakisConsistentSpectralApproximation2024}. This metric measures the distance of the eigenfunction $\zeta_k$ from a perfectly oscillatory eigenfunction with frequency $\omega_k$; a smaller score is better. Second, we sort by the Dirichlet energy of each eigenfunction, which measures eigenfunction regularity; a lower energy indicates a more regular (better) eigenfunction. This serves as a proxy for numerical approximation error which has been used with empirical success in works including \citeA{dasReproducingKernelHilbert2021, lintnerIdentificationMaddenJulian2023, valvaQBOAnnualCycle2025}.  We precisely define these metrics in \ref{sec:eig_sort_appendix}.

Only two of the leading leading eigenpairs (marked by stars) in the intraseasonal range have favorable rankings under both metrics. To distinguish them, we examine their seasonality and propagation characteristics. One eigenfunction which we identify with the MJO (period $ \approx 50\, \mathrm{days}$) is more active during Boreal winter and has eastward-propagating variability. The other eigenmode (period $ \approx 53\, \mathrm{days}$) is more active during Boreal summer and exhibits meridional propagation, appearing to be associated with the Boreal Summer Intraseasonal Oscillation (BSISO). We leave analysis of this possible BSISO mode to future work, and provide figures demonstrating the seasonality and propagation of this mode in the supplement. Here, we focus on the mode that is more active during Boreal winter, which we will call the MJO Koopman index.

\section{The Koopman index captures key features of MJO variability}
\label{sec:comparison big}

\label{sec:hovs}
\begin{figure*}[h!]
    \centering
    \includegraphics[width=1\textwidth]{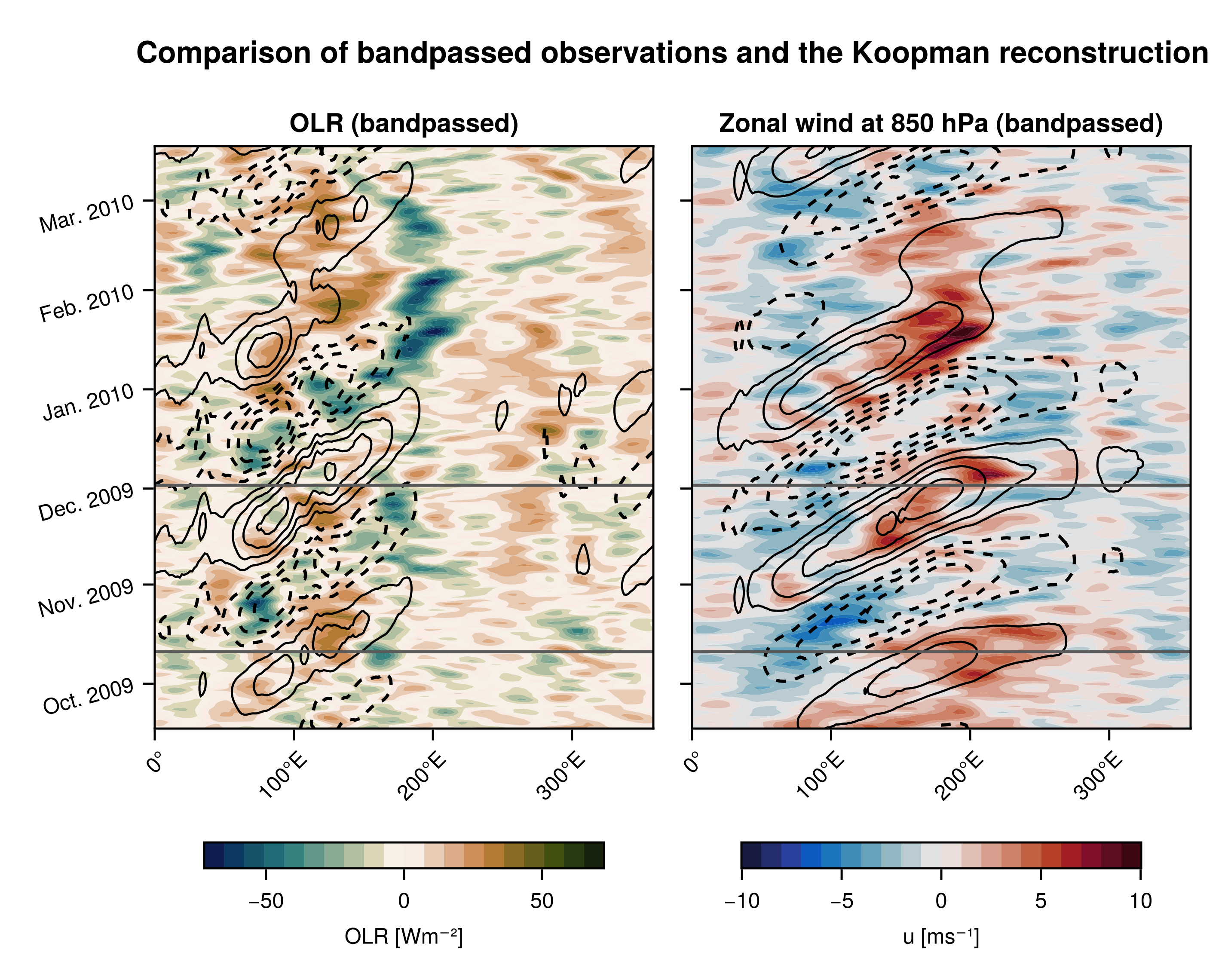}
    \caption{Intraseasonally bandpassed (20-96 days) OLR and zonal wind (filled contours) overlaid with the projection of the Koopman MJO (solid lines indicate positive values, dashed lines indicate negative values) during the YOTC period. Note the onset of two MJO events, seen by the initial band of negative OLR anomaly (enhanced convection): the first initiating in late October, the second in mid December. Approximate initiation dates (October 25 and 16 December) are marked by lines on the figure. The Koopman projection tracks the large scale location and propagation of these events.}
    \label{fig:hov}
\end{figure*}

To establish the fitness of the Koopman index for characterizing and predicting the MJO, we ask whether the mode captures the evolution of MJO events. We first consider the behavior of the Koopman mode as compared to intraseasonally bandpassed OLR and zonal wind during the Year of Tropical Convection (YOTC) virtual experiment \cite{waliserYearTropicalConvection2012}, focusing on the period of October 2009 to March 2010. 

Following past work \cite<including>{wheelerAllSeasonRealTimeMultivariate2004}, we bandpass OLR and zonal wind fields to retain variability between 20 and 96 days. Figure \ref{fig:hov} overlays contours of the Koopman MJO mode on this bandpassed data. In the YOTC period, there were two MJO initiation events: the first in late October and the second in mid-December, which can be seen in the figure as negative OLR anomalies (indicating regions of enhanced convection) that propagate eastward at about $5 \, \text{ms}^{-1}$ followed by similar propagation of positive OLR anomalies (as interpreted in \citeA{kiladisComparisonOLRCirculationBased2014}). These events are mirrored in the zonal wind with propagating westerlies followed by easterlies. The Koopman mode projected onto this data captures the initiation and propagation of both these events on the large scale, including the speed and location.

\label{sec:compare RMM}
\begin{figure*}[h!]
    \centering
    \includegraphics[width=\textwidth]{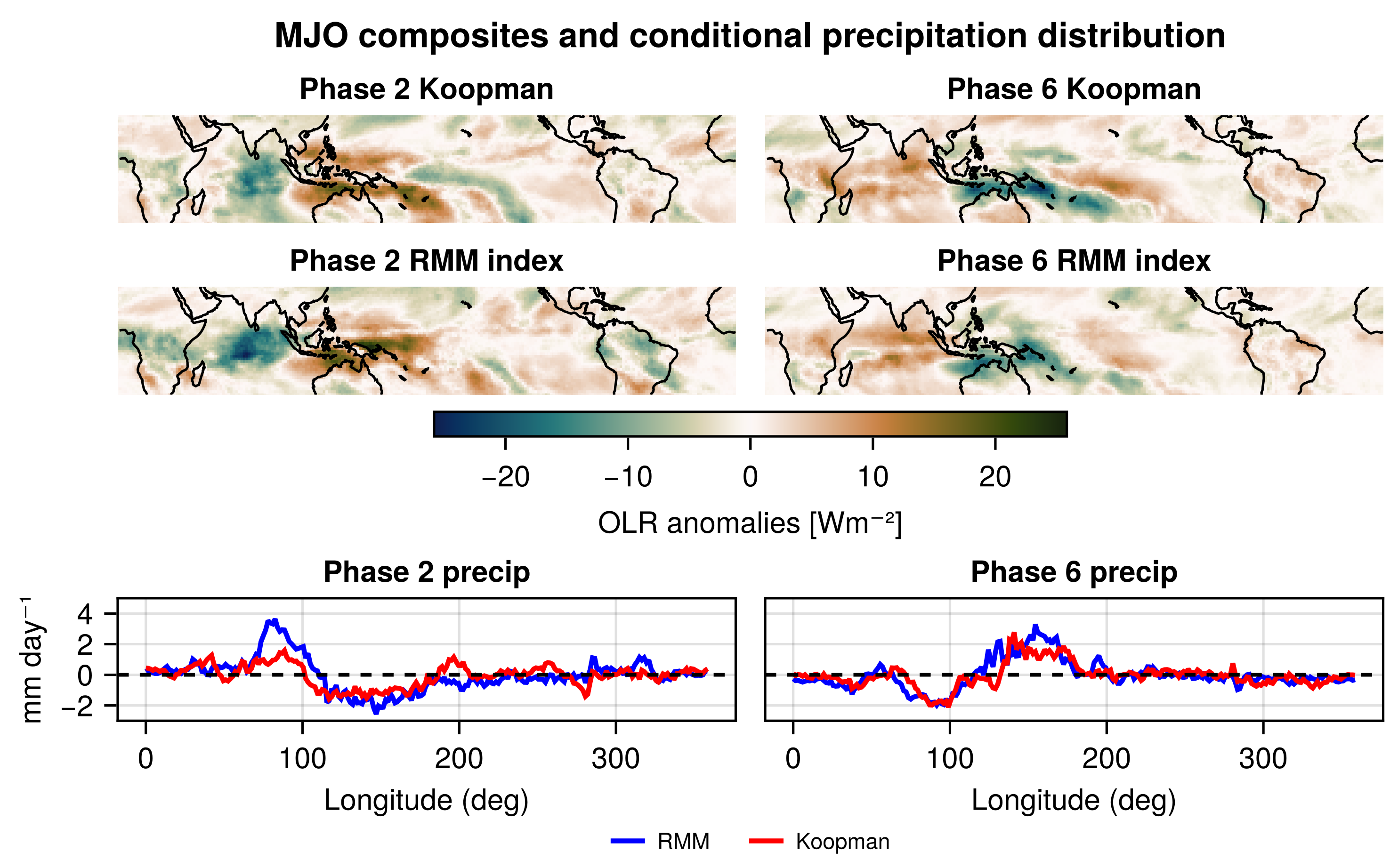}
    \caption{Comparison of the RMM and Koopman MJO representations. The top panel compares OLR composites from each of these indices in two phases. The bottom panel compares the conditional mean precipitation anomalies between the two indices. To focus on active MJO periods, we plot the precipitation anomalies (meridionally averaged between $15^\circ$ N - $15^\circ$ S) associated with times when the amplitude of the MJO index exceeds the 75th percentile of the respective index.}
    \label{fig:composite_comp}
\end{figure*}

We next examine the large-scale circulation and precipitation patterns associated with the Koopman index, comparing it with the RMM. We focus only on the extended Boreal Winter (DJFMA), as the selected Koopman index represents a boreal winter intraseasonal mode. The RMM is used as an all-season MJO index, but its boreal-summer variability also reflects aspects of the BSISO \cite{wheelerAllSeasonRealTimeMultivariate2004}.

So that we can directly compare the MJO phases between the two indices, we rotate the Koopman index to maximize the all-season correlation between the Koopman and RMM index, allowable as the complex phase of a Koopman eigenfunction is arbitrary. The DJFMA correlation between these two indices is $0.51$ indicating that the indices have related, but not identical evolution. (The correlation can be improved somewhat if we intraseasonally bandpass the RMM, increasing to $0.58$.) In the upper panel of figure \ref{fig:composite_comp} we plot the OLR composites for phases 2 and 6. The Koopman and RMM OLR composites show consistent representations of the MJO. Both exhibit similar large scale patterns of enhanced and suppressed convection that propagate in the same manner.

To reveal how these indices capture precipitation, we also show daily ERA5 precipitation anomalies associated with active MJO periods in phases 2 and 6 for the two respective indices in figure \ref{fig:composite_comp}. Specifically, we plot the precipitation anomaly conditioned on an index exceeding a $75\text{th}$ percentile threshold to ensure that the index is identifying an active MJO. Both indices identify a positive precipitation anomaly over the Indian Ocean and a negative anomaly over the Western Pacific in phase 2. In phase 6, this pattern is reversed. The spatial patterns of positive and negative daily precipitation anomalies are quite similar, although the Koopman index exhibits slightly weaker amplitude. We conclude that the Koopman representation captures key features of MJO variability at both the event and composite levels, including its large-scale propagation and associated precipitation.

\section{The Koopman MJO index is highly predictable}
\label{sec:prediction big}
The Koopman formalism selects for spatiotemporal modes with approximately oscillatory behavior, yielding an index constructed to provide a more predictable representation of  MJO-related behavior. We demonstrate this by comparing forecasts of the Koopman and RMM indices.

\begin{figure*}[h!]
    \centering
    \includegraphics[width=\textwidth]{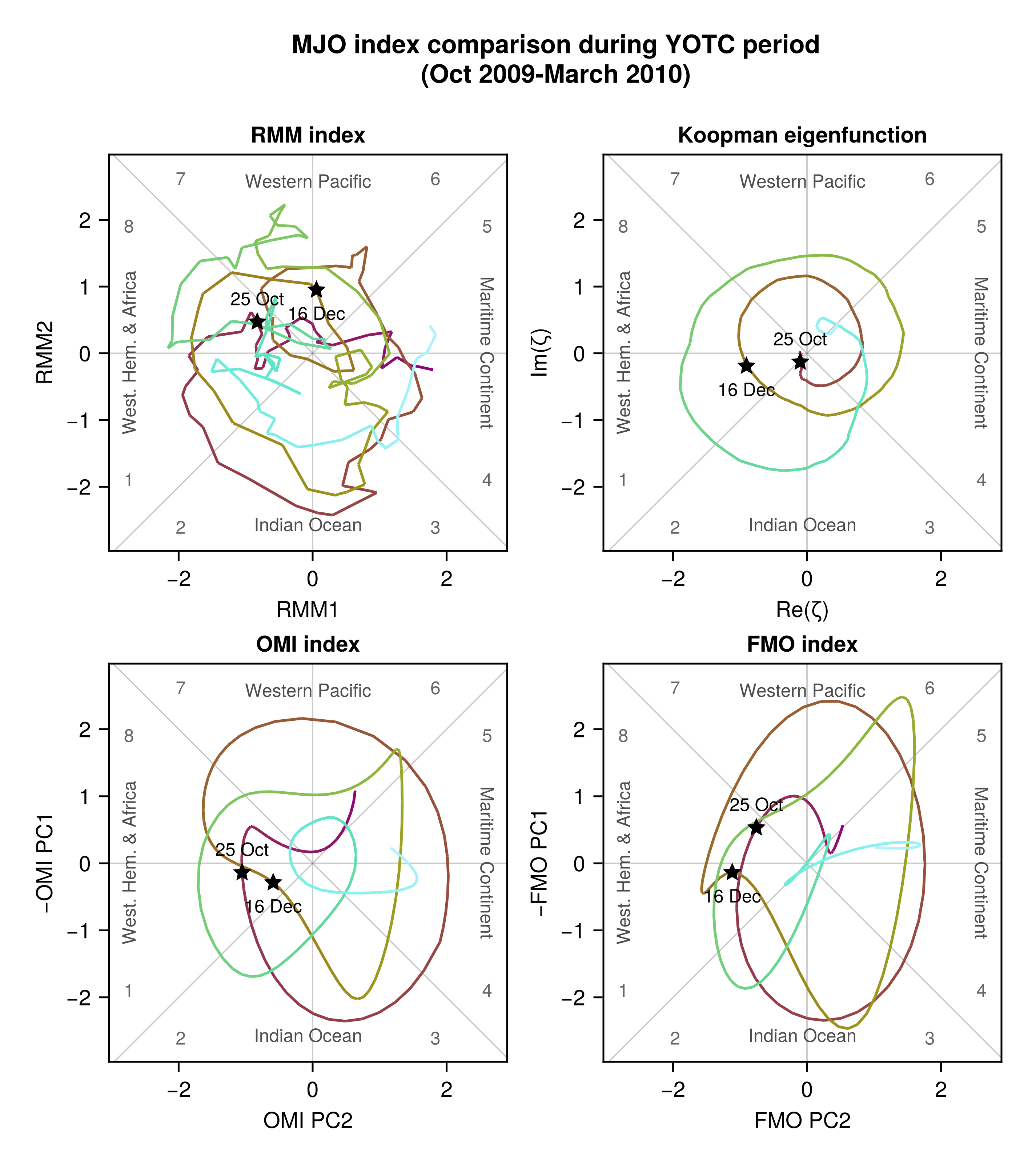}
    \caption{Comparison of the RMM, OMI, FMO and Koopman eigenfunctions during the YOTC period. Indices evolve (approximately) counterclockwise, with the color (red-to-blue) changing over time. The initiation dates of the two MJO events during this period (October $25\text{th}$, December $16\text{th}$) are marked by stars.}
    \label{fig:eig_comp}
\end{figure*}

While the Koopman and RMM indices capture similar MJO variability on large spatiotemporal scales, their short term evolution differs: the Koopman index propagates more smoothly than the RMM. In figure \ref{fig:eig_comp}, we plot the evolution of the Koopman MJO index as compared to the RMM during the YOTC period. While both track the counterclockwise propagation of the MJO, the evolution of the Koopman MJO index is much smoother and easier to interpret. The Koopman MJO exhibits two complete counterclockwise ``loops'' of the index in the complex plane, as well as a partial loop, which correspond to the two eastward propagating MJO events in this period and the initiation of the enhanced convection near the end of the YOTC period (Figure \ref{fig:hov}). While the RMM propagates counterclockwise as well, it sketches a fairly noisy, rough pattern. One can make out 3-4 rotations, depending on interpretation of the noise. This suggests that the RMM may be potentially harder to predict. 

The lower panels of Figure \ref{fig:eig_comp} allow comparison between the Koopman index and smoother MJO indices constructed from filtered data: the OMI (the OLR MJO Index) and the FMO (the Filtered OLR MJO index) \cite{kiladisComparisonOLRCirculationBased2014}. These indices evolve more smoothly like the Koopman MJO, but they are not “real-time” indices (i.e., cannot be computed instantaneously), and less useful for forecasting purposes.

We mark the two YOTC MJO initiation dates in the figure with stars. The Koopman index places both initiation dates in the same phase (1, corresponding to the Western Hemisphere/Africa quadrant), while the RMM and FMO separate the phase of these dates. The OMI and Koopman phases assign similar dynamical states to the initiation of two distinct MJO events. The RMM exhibits more inconsistency for the second event in particular, progressing into the corresponding phase later than the Koopman index. This aligns with previously reported differences between RMM phase progression and MJO convective evolution \cite{kiladisComparisonOLRCirculationBased2014}.

Finally, we establish that the smooth, consistent evolution of the Koopman index translates into forecast skill. We make forecasts from meridionally averaged ERA5 OLR and zonal wind using the same 64-day delay embedding as in the data analysis. To ensure a fair test, for a lead time $\ell$, the predictor consists of daily observations from $64 + \ell$ to $\ell$ days \textit{before} the forecast target. We evaluate forecast skill over DJFMA 2021--2024, defining the season based on the calendar year of January, which is completely distinct from the 2000--2019 analysis period that we used to establish the Koopman index.   

We use kernel analog forecasts (KAFs) to predict the Koopman and RMM indices. KAF is a nonparametric regression method in which predictions are constructed from the evolution of historical states, weighted according to their similarity to the current observation \cite{alexanderKernelAnalogForecasting2017}. The idea is rather simple: a forecast is made by finding similar conditions in the past and then using the associated history as the forecast. A kernel function measures the similarity. We use a reduced kernel eigenfunction basis, as described in \ref{sec:KAF}, and use only data from the analysis period.

\begin{figure*}[h]
    \centering
    \includegraphics[width=\textwidth]{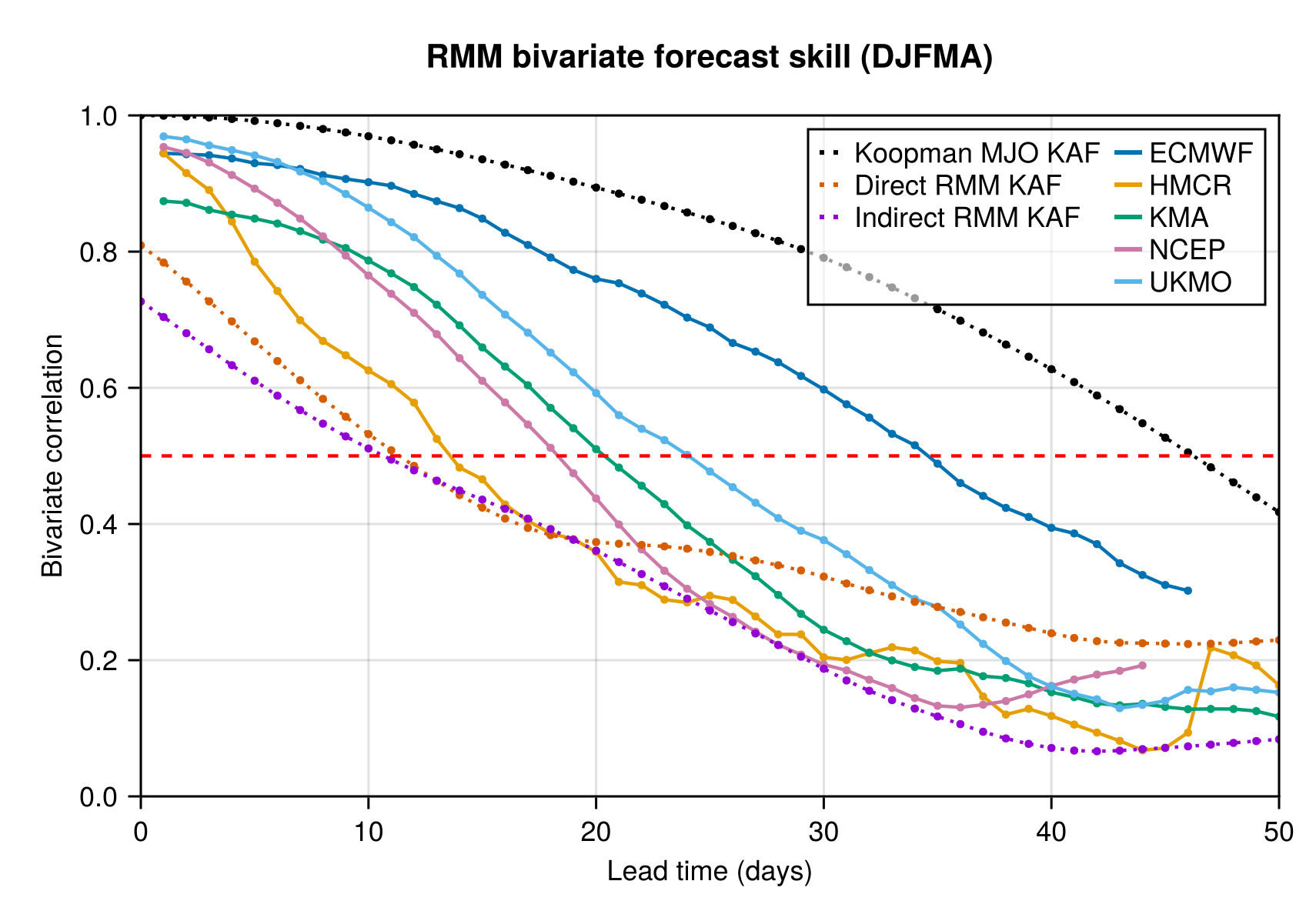}
    \caption{Comparison of S2S and KAF forecasts of RMM and the Koopman MJO Index. Solid lines are forecasts of the RMM from the S2S Prediction Project, while the dotted lines are KAF forecasts of the RMM and the Koopman MJO index. The KAF forecast of the Koopman MJO index is skillful for $46\, \textrm{days}$. In contrast, S2S forecasts of the RMM show skill ranging from $13$ to $35 \, \textrm{days}$, while the KAF forecast of RMM is only skillful for 11 days. }
    \label{fig:predict_comp}
\end{figure*}

We construct three KAF forecasts: direct forecasts of both the Koopman and RMM indices and an indirect RMM forecast obtained by first forecasting the Koopman index and then estimating RMM from the predicted Koopman state. For comparison, we also evaluate a subset of the operational S2S forecasts of the RMM from the S2S Prediction Project \cite{vitartSubseasonalSeasonalS2S2017a}. Figure \ref{fig:predict_comp} compares these forecasts using the bivariate correlation as a metric. We consider a correlation of 0.5 or above a skillful forecast. 

The smooth evolution of the Koopman MJO index makes it extremely predictable: simple analog forecasts maintain predictive skill for $46\, \textrm{days}$. This is well beyond the ability of the operational S2S models to forecast the MJO via the RMM, which are skillful for 13 to 35 days depending on the model \cite<consistent with S2S forecast skill reported in previous studies:>{limMJOPredictionSkill2018, zhouUnderstandingFactorsControlling2024a}.

The difference in forecasting skill lies primarily in that the S2S models are aiming for a harder target, the RMM, as compared to the smoothly evolving Koopman index. To illustrate this, we include the two KAF forecasts of RMM derived from the Koopman indices. In figure \ref{fig:predict_comp}, the dashed purple line shows how well we can forecast the RMM based on the Koopman MJO index alone: skillful forecasts are only possible for 11 days. The red dashed line shows a direct KAF forecast, but the improvement is minimal. The predictive skill of the RMM KAF forecasts decays at a similar rate to the S2S models, but begins with a rather low correlation value at day 1. The Koopman approach cannot capture the high-frequency variations of the RMM, as it selects for coherent, intraseasonal variability.

These results indicate that the Koopman index is substantially easier to forecast than the RMM. We emphasize that the Koopman and RMM indices do not precisely align. Nonetheless, the Koopman and RMM indices represent the same large scale variability associated with the MJO. Our main argument is that the Koopman index may be a better forecasting target in the subseasonal range.

\section{Discussion}
The Koopman approach provides a way to construct a data-driven representation of the MJO that is both physically interpretable and forecastable. The Koopman index captures primary features of the MJO variability, including its large-scale propagation and phase-dependent precipitation anomalies. Despite providing a similar MJO representation to the RMM, the two indices differ in their predictability. The Koopman index remains skillfully predictable up to 46 days with a simple kernel analog forecast compared to the RMM's 11. By comparison, the operational RMM forecasts remain skillful to 35 days. 

The ability to retain MJO characteristics while producing a more predictable index suggests that Koopman-based representations may provide useful alternative targets for extended-range forecasting. In addition, the construction of Koopman indices can be adapted to specific forecasting objectives. Here, we have chosen OLR and zonal winds as our input data to allow for direct comparison with the RMM. Including precipitation anomalies directly in the initial analysis step, however, could produce Koopman modes that more explicitly represent the precipitation variability associated with the MJO.

\section*{Open Research Statement}
ERA5 zonal wind, outgoing longwave radiation, and precipitation data were obtained from the ECMWF Climate Data Store \cite{hersbachERA5GlobalReanalysis2020}, associated with DOI \texttt{10.24381/cds.bd0915c6} and  \texttt{10.24381/cds.adbb2d47}. 

RMM indices from the S2S Prediction Project forecasts were obtained from
\url{https://confluence.ecmwf.int/spaces/S2S/pages/40796953/Tools#Tools-MJORMMsindices}. We refer to \citeA{vitartSubseasonalSeasonalS2S2017a} for details of the S2S forecast database.

Observed RMM indices were obtained from the Australian Bureau of Meteorology:
\url{http://www.bom.gov.au/climate/mjo/#tabs=Monitoring}, \cite{bomRMM}.

The OMI and FMO values were obtained from \url{https://psl.noaa.gov/mjo/mjoindex/}, \cite{noaaindex}.

The code used for analysis and production of the figures can be found on GitHub at \url{https://github.com/clairevalva/predictableMJO} and is archived on Zenodo \cite{valvaPredictableMJO2026}.

\section*{AI Disclosure}
The Julia code used in analysis was developed with the assistance of Codex (GPT 5.5 and 5.6). Authors take full responsibility for accuracy.

\section*{Acknowledgments}
We thank Dimitrios Giannakis for helpful conversations on this work. E.G. and C.V. acknowledge support from NSF OAC-2004572 and Schmidt Sciences, as part of the Virtual Earth System Research Institute (VESRI). C.V.'s postdoctoral work, during which the manuscript was completed, is supported by the DoD Vannevar Bush Faculty Fellowship N00014-22-1-2790.


\appendix

\section{Kernel basis and kernel analog forecasting}
\label{sec:KAF}

Both the Koopman approximation and kernel analog forecasting used in this study employ a reduced basis constructed from eigenfunctions of a kernel integral operator. Let $\mathbf{X}_j \in \mathbb{R}^{N_d}$  denote the delay-embedded observations defined in Section~\ref{sec:theory data}, and let $k:\mathbb{R}^{N_d}\times\mathbb{R}^{N_d} \to \mathbb{R}$ denote the normalized kernel used in the analysis. The corresponding empirical kernel operator $\mathcal{K}_{N_t}$ acts on functions $g$ as
\begin{equation}
(\mathcal{K}_{N_t} g)(\mathbf{Y}) = \frac{1}{N_t}\sum_{j=0}^{N_t-1} k(\mathbf{Y},\mathbf{X}_j)\,g(\mathbf{X}_j),
\end{equation}

with eigenfunctions $\phi_j$ and eigenvalues $\lambda_j$ satisfying
\begin{equation}
\label{eq:kernel eig}
\mathcal{K}_{N_t} \phi_j = \lambda_j \phi_j.
\end{equation}

For a new datapoint $\mathbf{Y} \in \mathbb{R}^{N_d}$, an eigenfunction with $\lambda_j \neq 0$ can be evaluated using the Nystr\"om extension,
\begin{equation}
\phi_j(\mathbf{Y}) = \frac{1}{N\lambda_j} \sum_{n=0}^{N-1} k(\mathbf{Y},\mathbf{X}_n)\,\phi_j(\mathbf{X}_n).
\end{equation}
The value of $\phi_j(\mathbf{Y})$ is determined by a kernel-weighted average over training states, with larger contributions from states $\mathbf{X}_n$ that are close to $\mathbf{Y}$ in the kernel-induced geometry.

Kernel analog forecasting (KAF) uses this basis to estimate the future value of a forecast observable. Let $f:\mathbb{R}^{N_d}\to\mathbb{C}$ denote a scalar- or complex-valued observable, so that $f(\mathbf{X}_n)$ is its value at time $n$. For a forecast lead $\ell$, we approximate the map from a current observation $Y$ to the future value of the observable using the first $L$ kernel eigenfunctions,
\begin{equation}
\hat{f}_\ell(\mathbf{Y}) = \sum_{j=0}^{L-1} c_{j,\ell}\,\phi_j(\mathbf{Y}), \qquad c_{j,\ell} = \frac{1}{N_t-\ell} \sum_{n=0}^{N-\ell-1} \phi_j(\mathbf{X}_n)^{*}\,f(\mathbf{X}_{n+\ell}).
\end{equation}
Thus, $\hat{f}_\ell(\mathbf{Y})$ estimates the value of $f$ at lead $\ell$ from the current observation $Y$.

We use this construction for three forecasts. First, we directly forecast the  Koopman MJO index (with data $\mathbf{Z}_n \in \mathbb{C}$) and the RMM index (with data $\mathbf{R}_n = \mathrm{RMM1}_n + i\,\mathrm{RMM2}_n$). 

We also make an indirect forecast, where we first predict the Koopman index using KAF and then regress the RMM onto the Koopman index to estimate the RMM index. The direct and indirect RMM experiments distinguish between predicting the RMM field directly from the OLR and zonal wind fields and predicting it via the intermediate Koopman representation.

\section{Eigenvalue sorting}
\label{sec:eig_sort_appendix}

\subsection{Autocorrelation sorting}
\label{sec:ac_sorting}
We sort the eigenfunctions by the \textit{maximal autocorrelation error}, which measures the extent to which a computed eigenfunction $\zeta_k$ departs from ideal Koopman eigenfunction behavior over a specified time interval, providing an estimate of predictive skill.

Let $C_f(t)$ denote the time-autocorrelation function of a function $f$ with zero-mean and norm one, i.e. $C_f(t) = \langle f, K^t f \rangle$. If $\zeta$ is a perfect Koopman eigenfunction with eigenfrequency $\omega \in \mathbb{C}$ then  $C_\zeta(t) = e^{\omega t}$. This, as well as a desire to favor eigenfunctions whose autocorrelation remains coherent over the timescale of interest, motivates the use of $\epsilon_{T_c}(\omega, \zeta)$ as a sorting parameter
\begin{equation}
\label{eq:sortac}
    \epsilon_{T_c}(\omega, \zeta) = \max_{t \in [-T_c, T_c]} \operatorname{Re} (1 - e^{-\omega t} C_\zeta(t)), 
\end{equation}
where $T_c$ is the time window of interest (here, 45 days). A perfect Koopman eigenfunction-eigenvalue pair will have $\epsilon_{T_c}(\omega, \zeta) = 0$ and all others will have $\epsilon_{T_c}(\omega, \zeta) \in (0, 1]$. As such $\epsilon_{T_c}(\omega, \zeta)$ quantifies the departure of a function from Koopman eigenfunction behavior over the interval $[-T_c,T_c]$, and is closely related to the $\epsilon-$approximate spectrum of the Koopman operator \cite<see>{giannakisConsistentSpectralApproximation2024}.

\subsection{Dirichlet energy}
We also sort the computed Koopman eigenfunctions by their Dirichlet energy, a measure of eigenfunction regularity. Let $\lambda_k$ and $\phi_k$ denote the kernel eigenvalues and corresponding orthonormal eigenfunctions as defined in eq. \ref{eq:kernel eig}. Associated with each $\phi_k$ is the Dirichlet energy
\begin{equation}
    \eta_k
    =
    \frac{1}{\epsilon}
    \left(\frac{1}{\lambda_k}-1\right),
\end{equation}
where $\epsilon$ is the kernel bandwidth parameter. The kernel eigenvalues are ordered from largest to smallest, $\eta_k$ increases with $k$.

Each computed Koopman eigenfunction $\zeta_j$ is represented in this basis as
\begin{equation}
    \zeta_j
    =
    \sum_{k=0}^{L-1}
    c_{j,k}\phi_k.
\end{equation}
Because the $\phi_k$ are orthonormal eigenfunctions of the corresponding diffusion operator, the Dirichlet energy of $\zeta_j$ is
\begin{equation}
\label{eq:sortDE}
    E(\zeta_j)
    =
    \sum_{k=0}^{L-1}
    \eta_k |c_{j,k}|^2
    =
    \frac{1}{\epsilon}
    \sum_{k=0}^{L-1}
    \left(\frac{1}{\lambda_k}-1\right)
    |c_{j,k}|^2.
\end{equation}
Eigenfunctions whose expansion places more weight on high-energy kernel basis functions have larger Dirichlet energy and are less regular.

\end{document}